\documentclass[preprint,3p]{elsarticle} 
\usepackage{multicol,caption}

\newenvironment{Figure}
 {\par\medskip\noindent\minipage{\linewidth}}
 {\endminipage\par\medskip}

\usepackage{amssymb}

\usepackage{lineno}

\usepackage{graphicx}
\usepackage{dcolumn}
\usepackage{bm,color}
\usepackage{epsfig}
\usepackage{rotating}
\usepackage{float}
\usepackage{amsmath}

\journal{Physics Letters B}

\begin{document}

\begin{frontmatter}



\title{Dark Photon Search in the Mass Range Between
 \mbox{1.5 and 3.4 GeV/$c^2$}} 

\begin{small}
\author{
M.~Ablikim$^{1}$, M.~N.~Achasov$^{9,f}$, X.~C.~Ai$^{1}$, O.~Albayrak$^{5}$, M.~Albrecht$^{4}$, D.~J.~Ambrose$^{44}$, A.~Amoroso$^{48A,48C}$, F.~F.~An$^{1}$, Q.~An$^{45,a}$, J.~Z.~Bai$^{1}$
% ...
\\
\vspace{0.2cm}
(BESIII Collaboration)\\
\vspace{0.2cm} {\it
$^{1}$ Institute of High Energy Physics, Beijing 100049, People's Republic of China\\
% ...
$^{4}$ Bochum Ruhr-University, D-44780 Bochum, Germany\\
$^{5}$ Carnegie Mellon University, Pittsburgh, Pennsylvania 15213, USA\\
% ...
$^{9}$ G.I. Budker Institute of Nuclear Physics SB RAS (BINP), Novosibirsk 630090, Russia\\
% ...
$^{44}$ University of Rochester, Rochester, New York 14627, USA\\
$^{45}$ University of Science and Technology of China, Hefei 230026, People's Republic of China\\
% ...
$^{48}$ (A)University of Turin, I-10125, Turin, Italy; (B)University of Eastern Piedmont, I-15121, Alessandria, Italy; (C)INFN, I-10125, Turin, Italy\\
% ...
$^{a}$ Also at State Key Laboratory of Particle Detection and Electronics, Beijing 100049, Hefei 230026, People's Republic of China\\
% ...
$^{f}$ Also at the Novosibirsk State University, Novosibirsk, 630090, Russia\\
% ...
}}

\vspace{0.4cm}
\end{small}


\begin{abstract}
Using a data set of 2.93 fb$^{-1}$ taken at a center-of-mass energy
$\sqrt{s}$ = 3.773 GeV with the BESIII detector at the BEPCII collider, 
we perform a search for an extra U(1) gauge boson, also denoted as 
a dark photon. We examine the initial state radiation reactions 
$e^+e^-\rightarrow e^+e^-\gamma_{\rm ISR}$ and $e^+e^-\rightarrow 
\mu^+\mu^-\gamma_{\rm ISR}$ for this search, where the dark photon would 
appear as an enhancement in the invariant mass distribution of the 
leptonic pairs.  We observe no obvious enhancement in the mass 
range between 1.5 and 3.4 GeV/$c^{2}$ and set a 90\% confidence level
upper limit on the mixing strength of the dark photon and the Standard Model 
photon. We obtain a competitive limit in the tested mass range.
\end{abstract}

\end{frontmatter}




\begin{multicols}{2}
Several astrophysical anomalies, which cannot be easily
understood in the context of the Standard Model (SM) of particle physics 
or astrophysics, have been discussed in relation to a dark, so far 
unobserved sector~\cite{Arkani}, which couples very weakly with SM 
particles. The most straightforward model consists of an extra U(1) 
force carrier, also denoted as a dark photon, $\gamma'$, which couples 
to the SM via kinetic mixing~\cite{kinMix}. It has been shown in 
Ref.~\cite{Arkani} that the dark photon has to be relatively light, on 
the MeV/$c^{2}$ to GeV/$c^{2}$ mass scale, to explain the astrophysical
observations. Furthermore, it was realized, that a dark photon of 
similar mass could also explain the presently observed deviation on 
the level of 3 to $4\sigma$ between the measurement and the SM prediction 
of $(g-2)_\mu$~\cite{Pospelov}. These facts and the work by Bjorken 
and collaborators~\cite{darkPhoton_bjorken} triggered  
searches for the dark photon at particle accelerators in a world wide 
effort~\cite{Batell,Haibo}. Different experimental setups can be used, 
like fixed-target (e.g. Refs.~\cite{darkPhoton_MAMI2, darkPhoton_MAMI}), 
beam dump (e.g. Refs.~\cite{darkPhoton_E774, darkPhoton_E141}), or 
low-energy collider experiments (e.g. Refs.~\cite{darkPhoton_BABAR1,darkPhoton_BABAR_new}). 
The mixing strength $\varepsilon = \alpha' /\alpha$, where $\alpha'$ is 
the coupling of the dark photon to the electromagnetic charge and $\alpha$ 
the fine structure constant, is constrained by previous measurements to 
be below approximately~$10^{-2}$~\cite{darkPhoton_bjorken}.
\\

In this letter we present a dark photon search, using 2.93 fb$^{-1}$~\cite{2pi_BES} 
of data taken at $\sqrt{s}$ = 3.773 GeV obtained with the Beijing 
Spectrometer III (BESIII). The measurement exploits the process of 
initial state radiation (ISR), in which one of the beam particles 
radiates a photon. In this way, the available energy to produce final 
states is reduced, and the di-lepton invariant masses below the 
center-of-mass energy of the $e^+e^-$ collider become available. The 
same method has been used by the BaBar experiment~\cite{darkPhoton_BABAR1,darkPhoton_BABAR_new}, 
where a dark photon mass $m_{\gamma^\prime}$ between 0.02 and 10.2
GeV/$c^{2}$ and $\varepsilon$ values in the order of 10$^{-3}$ - 10$^{-4}$ 
have been excluded. We search for the processes $e^+e^-\rightarrow
\gamma' \gamma_{\rm ISR} \rightarrow l^+l^-\gamma_{\rm ISR}$ ($l=\mu ,e$) with 
leptonic invariant masses $m_{l^{+}l^{-}}$ between 1.5 and 3.4 GeV/$c^{2}$. 
The ISR QED processes $e^+e^-\rightarrow \mu^+\mu^-\gamma_{\rm ISR}$ and 
$e^+e^-\rightarrow e^+e^-\gamma_{\rm ISR}$ are irreducible background 
channels. However, the dark photon width is expected to be smaller than 
the resolution of the experiment~\cite{darkPhoton_bjorken} and, thus, 
a $\gamma'$ signal would lead to a narrow structure at the mass
of the dark photon in the $m_{l^{+}l^{-}}$ mass spectrum on top of the 
continuum QED background. \\

The BESIII detector is located at the double-ring $e^+e^-$ Beijing
Electron Positron Collider (BEPCII)~\cite{BESIII}.  The cylindrical
BESIII detector covers 93\% of the full solid angle. It consists of
the following detector systems.  (1) A Multilayer Drift Chamber (MDC)
filled with a helium-gas mixture, composed of 43 layers, which provides 
a spatial resolution of 135 $\mu$m and a momentum resolution of 0.5\% 
for charged tracks at 1 GeV/$c$ in a magnetic field of 1 T.  (2) A
Time-of-Flight system (TOF), built with 176 plastic scintillator
counters in the barrel part, and 96 counters in the end caps. The time
resolution in the barrel (end caps) is 80 ps (110 ps). For momenta 
up to 1~GeV/$c$, this provides a 2$\sigma$ K/$\pi$ separation.
(3) A CsI(Tl) Electro-Magnetic Calorimeter (EMC) with an energy
resolution of 2.5\% in the barrel and 5\% in the end caps at an energy
of 1 GeV.  (4) A Muon Counter (MUC) consisting of nine barrel and
eight endcap resistive plate chamber layers with a 2 cm position
\mbox{resolution.}  \\


For the simulation of ISR processes $e^+e^-\rightarrow
\mu^+\mu^-\gamma_{\rm ISR}$ and $\pi^+\pi^-\gamma_{\rm ISR}$, the {\sc phokhara} 
event generator~\cite{Phokhara,Phokhara7}, which includes ISR and final 
state radiation (FSR) corrections up to next-to-leading order, is used. 
Bhabha scattering is simulated with {\sc babayaga 3.5}~\cite{BABAYAGA}. 
Continuum Monte Carlo (MC) events, as well as the resonant $\psi (3770)$ 
decays to $D\bar{D}$, non-$D\bar{D}$, and the ISR production of $\psi^\prime$ 
and $J/\psi$, are simulated with the {\sc kkmc} generator~\cite{KKMC}. 
All MC generators, which are the most appropriate choices for the processes 
studied, have been interfaced with the {\sc geant4}-based~\cite{GEANT1,GEANT2} 
detector simulation.
\\ 

The selection of $\mu^+\mu^-\gamma_{\rm ISR}$ and $e^+e^-\gamma_{\rm ISR}$
events is straightforward. We require the presence of two charged
tracks in the MDC with net charge zero. The points of closest approach
from the interaction point (IP) for these two tracks are required to
be within a cylinder of 1 cm radius in the transverse direction and
$\pm$10 cm of length along the beam axis.  The polar angle with 
respect to the beam axis $\theta$ of the tracks is required to be in 
the fiducial volume of the MDC: 0.4 $< \theta < \pi-0.4$ 
radians. In order to suppress spiraling tracks, we require the transverse 
momentum $p_t$ to be above 300~MeV/$c$ for both tracks.

Muon particle identification is used~\cite{BESIII_2}. The
probabilities for being a muon $P(\mu)$ and being an electron $P(e)$
are calculated using information from MDC, TOF, EMC, and MUC. 
For both charged tracks, $P(\mu)>P(e)$ is required.
To select electrons, the ratio of the measured energy
in the EMC, $E$, to the momentum $p$ obtained from the MDC is
used. Both charged tracks must satisfy $E/p$ $>$ 0.8~$c$.

\begin{figure*}[htp]
   \centering 
   \includegraphics[width=3.0 in, height=2.0in]{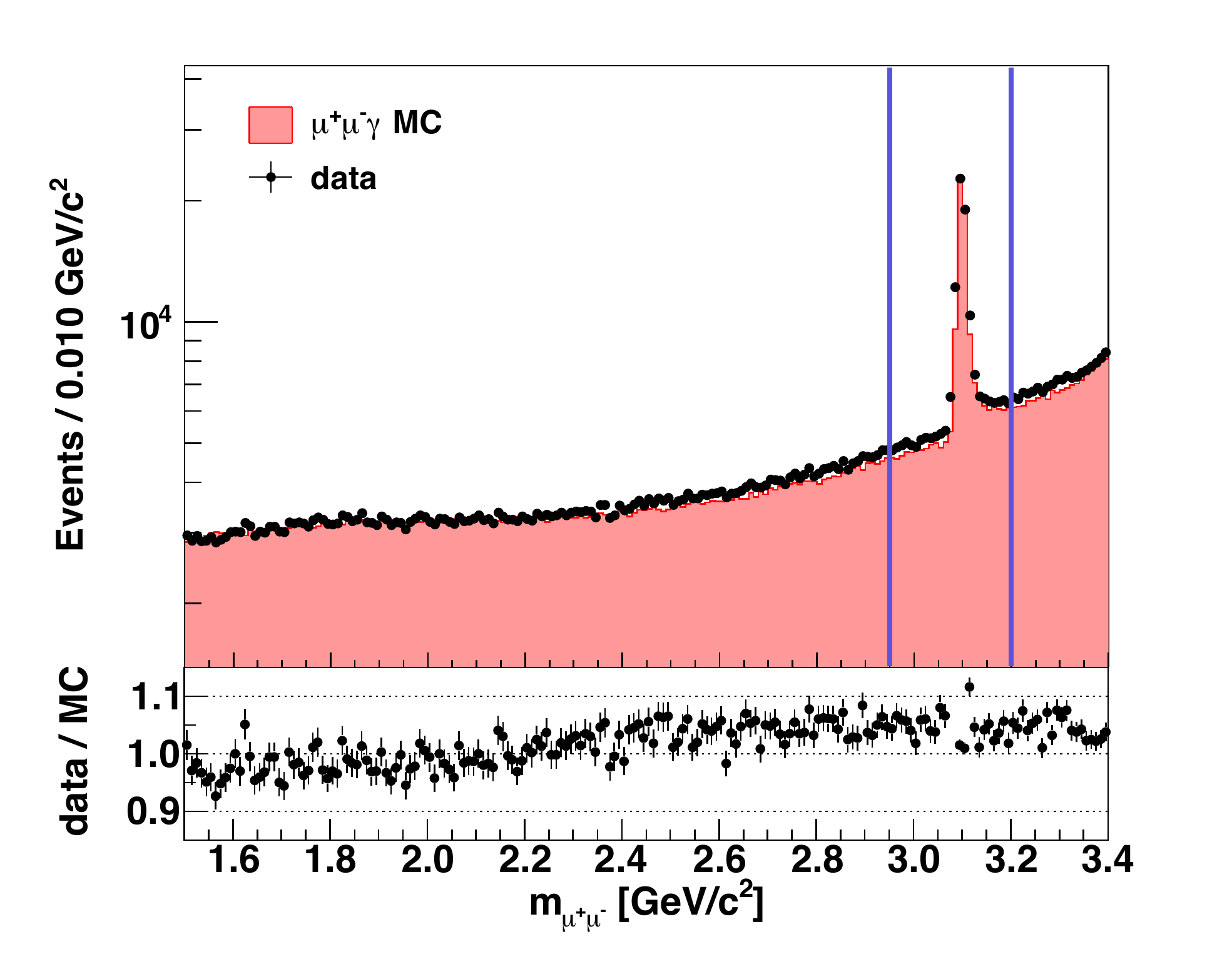} 
   \includegraphics[width=3.0 in, height=2.0in]{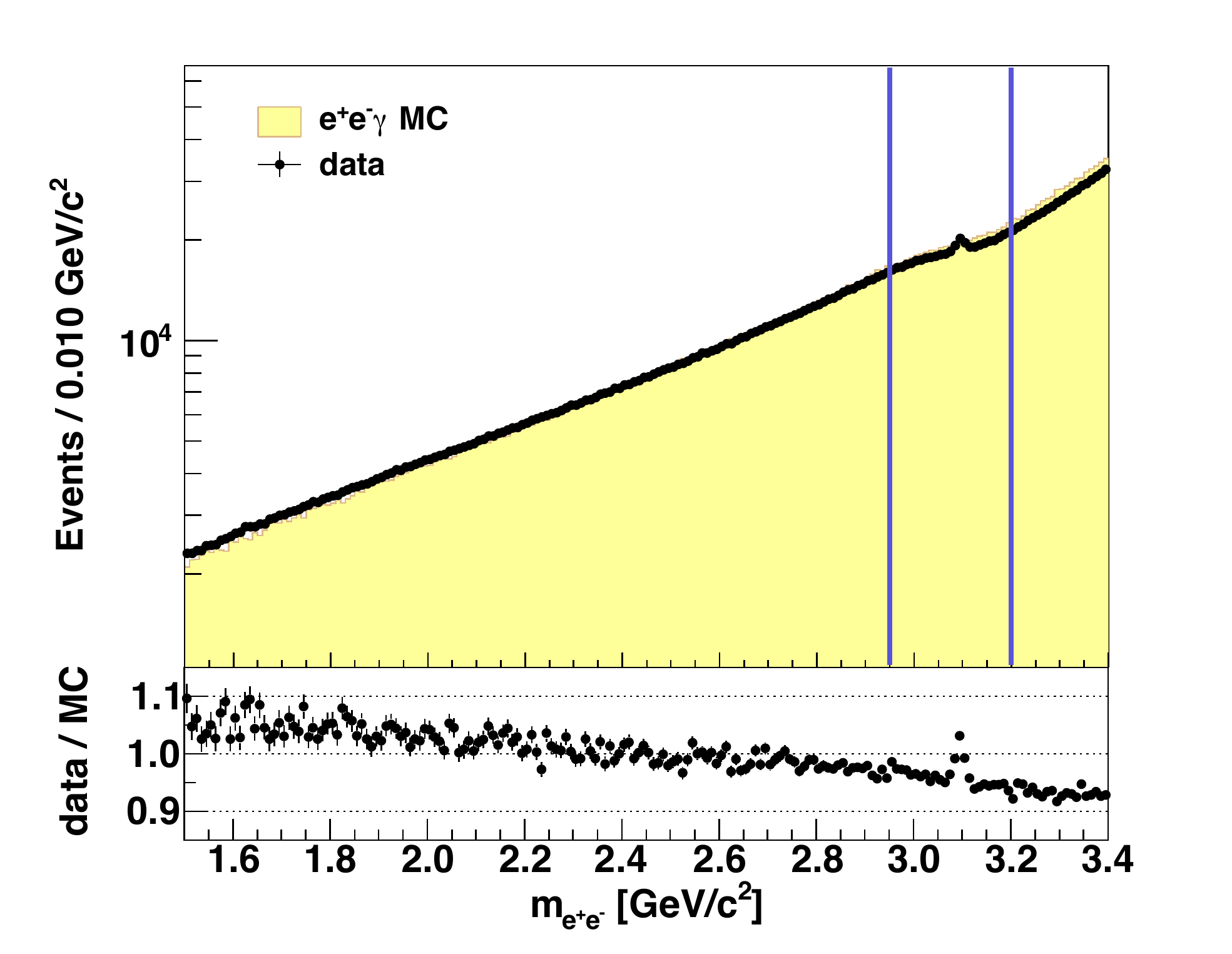}
   \captionof{figure}{Leptonic invariant mass distributions $m_{\mu^{+}\mu^{-}}$
   and $m_{e^{+}e^{-}}$ after applying the selection requirements. Shown is
   data (points) and MC simulation (shaded area), which is scaled to the
   luminosity of the data set. The marked area around the $J/\psi$
   resonance is excluded in the analysis. 
   The lower panel shows the ratio of data and MC simulation (points) and
   the ratio of fit curve and MC simulation (histogram).}
   \label{invMass_muon}
\end{figure*}

The radiator function~\cite{radiator function}, which describes the 
radiation of an ISR photon, is peaked at small $\theta$ values with 
respect to the beam axis. Different from BaBar, we use untagged ISR 
events, where the ISR photon is emitted at a small angle $\theta_\gamma$ 
and is not detected within the angular acceptance of the EMC, to increase 
statistics.
A one constraint (1C) kinematic fit, applying energy and momentum
conservation, is performed with the hypothesis
$e^+e^-\rightarrow\mu^+\mu^-\gamma_{\rm ISR}$ or $e^+e^-\rightarrow
e^+e^-\gamma_{\rm ISR}$, using as input the two selected charged track
candidates, as well as the four momentum of the initial $e^+e^-$
system. The constraint is the mass of a missing photon. The fit
quality condition $\chi^2_{1\rm C}$/{\rm (dof=1)} $<$ 20 is applied in the
$\mu^+\mu^-\gamma_{\rm ISR}$ case, where dof is the degree of freedom. 
To suppress non-ISR
background, the angle of the missing photon, $\theta_\gamma$,
predicted by the 1C kinematic fit, is required to be smaller than 0.1
radians or greater than $\pi - 0.1$ radians. We
apply stronger requirements for the $e^+e^-\gamma_{\rm ISR}$ final state,
to provide a better suppression of the non-ISR background
which is higher in the $e^{+}e^{-}$ channel compared to the 
$\mu^{+}\mu^{-}$ channel. 
In this case, \mbox{$\chi^2_{1\rm C}$/{\rm (dof=1)} $<$ 5}, and $\theta_\gamma
< 0.05$ radians, or \mbox{$\theta_\gamma >  \pi-0.05$ radians.}  \\

Background in addition to the radiative QED processes
$\mu^+\mu^-\gamma_{\rm ISR}$ and $e^+e^-\gamma_{\rm ISR}$, which is
irreducible, is studied with MC simulations and is negligible for the
$e^+e^-\gamma_{\rm ISR}$ final state, and on the order of 3\% for
$\mu^+\mu^-$ invariant masses below 2~GeV/$c^{2}$ due to muon
misidentification, and negligible above. This remaining background
comes mostly from $\pi^+\pi^-\gamma_{\rm ISR}$ events. We subtract their
contribution using a MC sample, produced with the {\sc phokhara}
generator. The subtraction of this background leads to a systematic
uncertainty due to the generator precision
smaller than 0.5\%. \\

The $\mu^+\mu^-$ and $e^+e^-$ invariant mass distributions,
$m_{\mu^{+}\mu^{-}}$ and $m_{e^{+}e^{-}}$, which are
shown separately in Fig.~\ref{invMass_muon}, are mainly dominated by
the QED background but could contain the signal sitting on top of
these irreducible events. For comparison with data, MC simulation,
scaled to the luminosity of data, is shown, although it is not used in
the search for the dark photon.  In this analysis, the dark photon
mass range $m_{\gamma'}$ between 1.5 and 3.4 GeV/$c^{2}$ is studied.
Below 1.5 GeV/$c^{2}$ the $\pi^+\pi^-\gamma_{\rm ISR}$ cross section with
muon misidentification dominates the $m_{\mu^{+}\mu^{-}}$
spectrum. Above 3.4~GeV/$c^{2}$ the hadronic $q\bar q$ process can not
be suppressed sufficiently by the $\chi^2_{1\rm C}$ requirement.  In
order to search for narrow structures on top of the QED
background, 4th order polynomial functions to describe the continuum QED are
fitted to the data distributions shown in Fig.~\ref{invMass_muon}.
The mass range around the narrow $J/\psi$ resonance between 2.95 and
3.2 GeV/$c^{2}$ is excluded.

The differences between the $\mu^+\mu^-\gamma_{\rm ISR}$ and
$e^+e^-\gamma_{\rm ISR}$ event yields and their respective 4th order
polynomials are added.
The combined differences are represented by the black dots in
Fig.~\ref{rolke}.  A dark photon candidate would appear as a peak in
this plot.  The observed statistical significances are less than
3$\sigma$ everywhere in the explored region. The significance in each
invariant mass bin is defined as the combined differences between data
and the 4th order polynomials, divided by the combined statistical
errors of both final states. In conclusion, we observe no dark photon
signal for 1.5 GeV/$c^{2}$ $<$ $m_{\gamma'}$ $<$ 3.4 GeV/$c^{2}$, where
$m_{\gamma'}$ is equal to the leptonic invariant mass
$m_{l^{+}l^{-}}$.
The exclusion limit at the 90\% confidence level is determined with a
profile likelihood approach~\cite{rolke}. Also shown in
Fig.~\ref{rolke} as a function of $m_{l^{+}l^{-}}$ is the
bin-by-bin calculated exclusion limit, including the systematic
uncertainties as explained below. 

\begin{Figure}
   \centering \includegraphics[width=2.5in, height=2.0in]{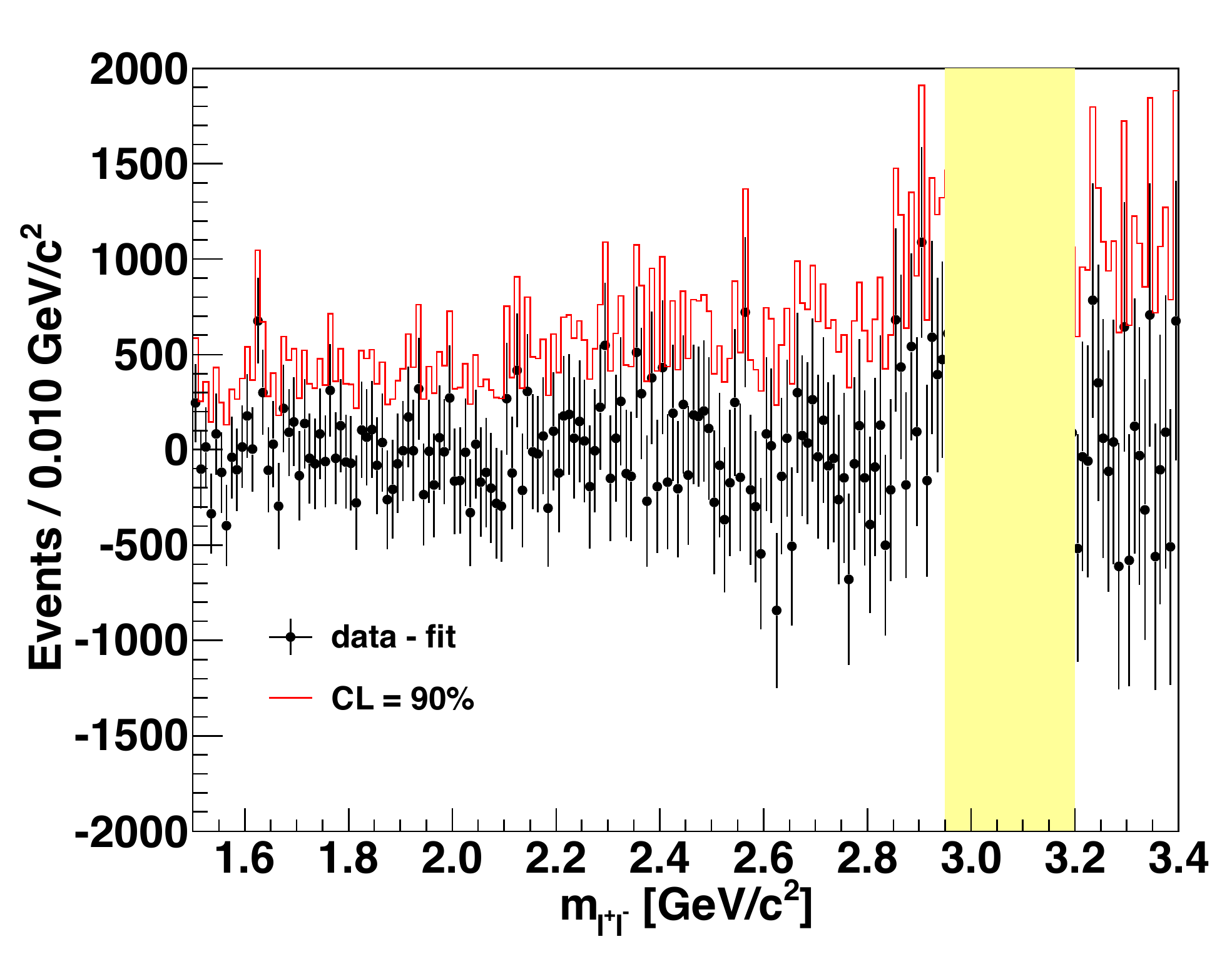}
   \captionof{figure}{The sum of the differences between the
     $\mu^+\mu^-\gamma_{\rm ISR}$ and $e^+e^-\gamma_{\rm ISR}$ event yields
     and their respective 4th order polynomials (dots with error
     bars).  The solid histogram represents the exclusion limit
     with the 90\% confidence, calculated with a profile likelihood
     approach and including the systematic uncertainty. The region
     around the $J/\psi$ resonance between 2.95 and 3.2~GeV/$c^{2}$ is
     excluded.}
   \label{rolke} 
\end{Figure}

To calculate the exclusion limit on
the mixing parameter $\varepsilon^2$, the formula from
Ref.~\cite{darkPhoton_bjorken} is used
\begin{linenomath}
\begin{equation}
\frac{\sigma_{i}(e^+e^- \rightarrow \gamma^\prime\,\gamma_{\rm ISR}\rightarrow l^+l^-\gamma_{\rm ISR})}{\sigma_i(e^+e^-\rightarrow \gamma^*\,\gamma_{\rm ISR}\rightarrow l^+l^-\gamma_{\rm ISR})} =  \nonumber  
\end{equation}
\begin{equation}
\frac{N^{\rm up}_{i}(e^+e^- \rightarrow \gamma^\prime\,\gamma_{\rm ISR}\rightarrow l^+l^-\gamma_{\rm ISR})}{N^{\rm B}_i(e^+e^-\rightarrow \gamma^*\,\gamma_{\rm ISR}\rightarrow l^+l^-\gamma_{\rm ISR})}\cdot\frac{1}{\epsilon} = \nonumber 
\end{equation}
\begin{equation}
\label{formula_excl_2mu}
\frac{3\pi\cdot\varepsilon^2 \cdot m_{\gamma'}}{2N_f^{l^+l^-}\alpha\cdot\delta_m^{l^+l^-}},
\end{equation}
\end{linenomath}
where $i$ represents the $i$-th mass bin, $\alpha$ is the
electromagnetic fine structure constant, $m_{\gamma'}$ the dark photon
mass, $\gamma^*$ the SM photon, and $\delta_m^{l^{+}l^{-}}$
($l=\mu,~e$) the bin width of the lepton pair invariant mass
spectrum, 10 MeV/$c^2$. The mass resolution of the lepton pairs
determined with MC for $e^{+}e^{-}$ and $\mu^{+}\mu^{-}$ is between
5 and 12 MeV/$c^{2}$.  The cross section ratio upper limit in
Eq.~\ref{formula_excl_2mu} is determined from the exclusion upper
limit ($N^{\rm up}$) corrected by the efficiency loss 
($\epsilon$) due to the bin width divided
by the number of $\mu^+\mu^-\gamma_{\rm ISR}$ and $e^+e^-\gamma_{\rm ISR}$
events ($N^{\rm B}$) corrected as described below.  
The efficiency loss caused by the incompleteness of signal events in 
one bin is calculated with $\int_{-5~{\rm MeV/}c^2}^{5~{\rm MeV/}c^2}\mathrm{G}(0,\sigma)\,
dm/\int_{-\infty}^{\infty}\mathrm{G}(0,\sigma)\, dm$, where
$\mathrm{G}(0,\sigma)$ is the Gaussian function used to describe the mass
resolution.

The QED cross section $\sigma_i(e^+e^-\rightarrow
\gamma^*\, \gamma_{\rm ISR}\rightarrow l^+l^-\gamma_{\rm ISR})$ must only take
into account annihilation processes of the initial $e^+e^-$ beam
particles, where a dark photon could be produced. Thus, the event
yield of the $e^+e^-\gamma$ final state has to be corrected
due to the existence of SM Bhabha scattering.  This
correction is obtained in bins of $m_{e^{+}e^{-}}$ by
dividing the $e^+e^-$ annihilation
events only by the sum of events of the annihilation and Bhabha
scattering processes. The first is generated with the {\sc phokhara}
event generator by generating the $\mu^+\mu^-\gamma$ final state and
replacing the muon mass with the electron mass. The latter is
generated with the {\sc babayaga@nlo} generator~\cite{BABAYAGA2}. The
correction factor varies between 2\% and 8\% depending on
$m_{e^{+}e^{-}}$.

The number of final states 
for the dark photon $N_f^{l^+l^-}$ includes the phase space above the $l^+l^-$ 
production threshold of the leptons $l = \mu , e$, and is given by $N_f^{l^+l^-} 
= \Gamma_{tot}/\Gamma_{ll}$~\cite{darkPhoton_tobias}, where $\Gamma_{ll} \equiv 
\Gamma(\gamma^\prime\rightarrow l^+l^-)$ is the leptonic $\gamma^\prime$ width 
and $\Gamma_{tot}$ is the total $\gamma'$ width. These widths are taken from 
Ref.~\cite{darkPhoton_tobias}
\begin{linenomath}
\begin{equation}
\Gamma_{ll} = \frac{\alpha\varepsilon^2}{3m_{\gamma^\prime}^2}(m_{\gamma^\prime}^2 + 2m_l^2)\sqrt{m_{\gamma^\prime}^2 - 4m_l^2}
\end{equation}
\begin{equation}
	\Gamma_{tot} = \Gamma_{ee}+ \Gamma_{\mu\mu}\cdot(1 + R(\sqrt{s})) \; ,
\end{equation}
\end{linenomath}
where $\Gamma_{ee} \equiv \Gamma(\gamma^\prime\rightarrow e^+e^-)$,
$\Gamma_{\mu\mu} \equiv \Gamma(\gamma^\prime\rightarrow \mu^+\mu^-)$,
and $R(\sqrt{s})$ is the total hadronic cross section $R$ 
value~\cite{PDG2014} as a function of $\sqrt{s}$.\\

The systematic uncertainties are included in the calculation of the
exclusion limit. The main source is the uncertainty of the $R$ value
taken from Ref.~\cite{PDG2014}, which enters the calculation of the
$N_f^{l^+l^-}$ and leads to a mass dependent systematic uncertainty
between 3.0 and 6.0\%. Other sources are background subtraction as
described above \mbox{($<$ 0.5\%)}, the fitting error of the
polynomial fit to data \mbox{($<$ 1\%)}, the Bhabha scattering
correction factor using the {\sc phokhara} and {\sc babayaga@nlo}
event generator ($<$ 1\%), and data-MC differences of the leptonic
mass resolution. To quantify the latter one, we study the data-MC
resolution difference of the $J/\psi$ resonance for the $\mu^+\mu^-$
and $e^+e^-$ decays, separately.  The resonance is fitted with a
double Gaussian function in data and MC simulation, and the width
difference is ($3.7\pm 1.8$)\% for $\mu^+\mu^-$ and
($0.7\pm 5.3$)\% for $e^+e^-$.  The differences are taken
into consideration in the calculations, and the uncertainty in the
differences (1\%) is taken as the systematic uncertainty of the
data-MC differences.
The mass dependent total systematic uncertainty, which varies from 3.5 to 6.5 \% 
depending on mass, is used bin-by-bin in the upper limit.  \\

\begin{figure*}
   \centering \includegraphics[width=10cm]{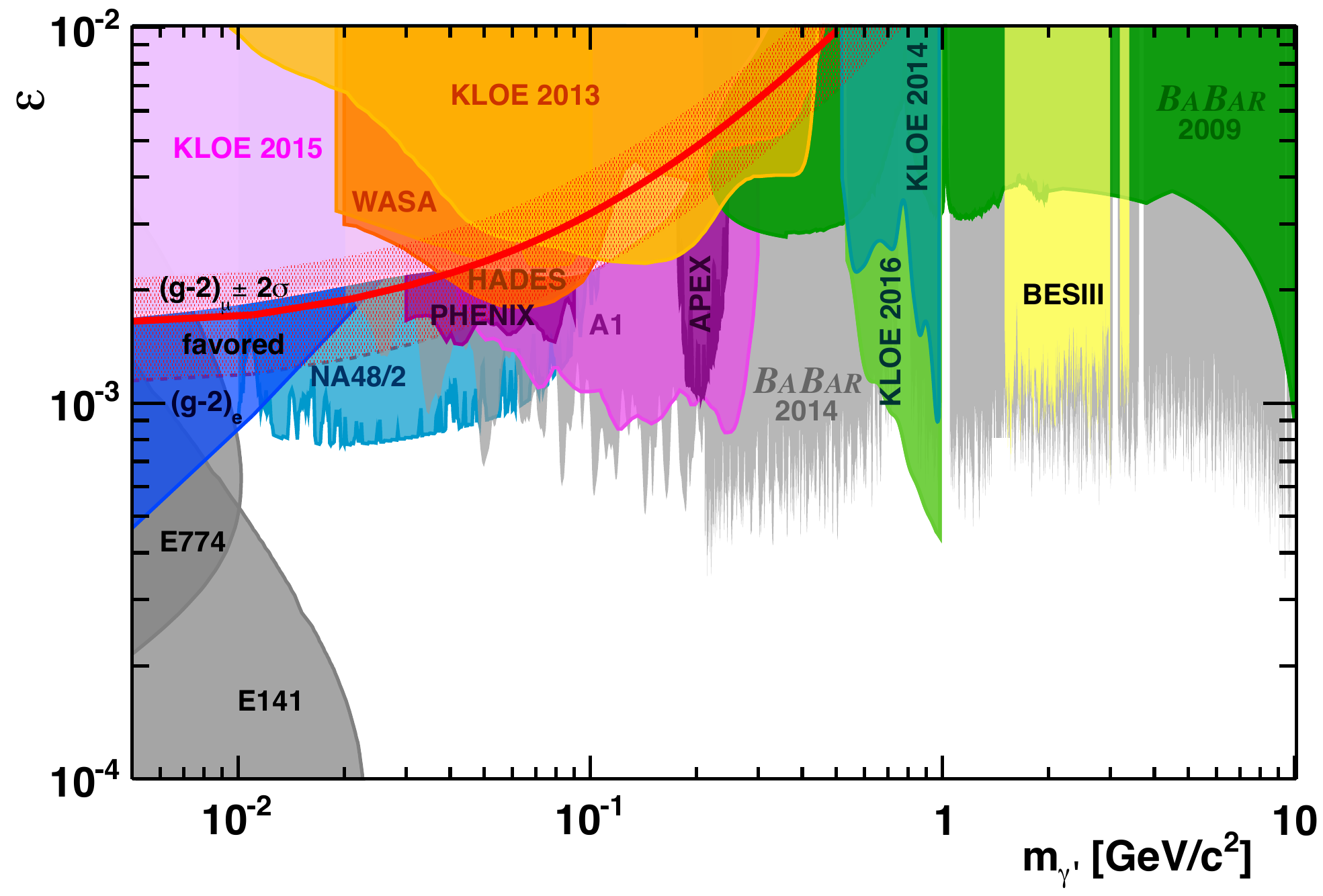}
   \captionof{figure}{Exclusion limit at the 90\% confidence level on the
   mixing parameter $\varepsilon$ as a function of the dark photon
   mass. The bold solid line represents the $\varepsilon$ values, which 
   would explain the discrepancy between the measurement and the SM 
   calculation of the anomalous magnetic moment of the muon~\cite{Pospelov}, 
   together with its 2$\sigma$ band.}
   \label{result}
\end{figure*}

The final result, the mixing strength $\varepsilon$ as a
function of the dark photon mass, is shown in Fig.~\ref{result},
including the systematic uncertainties. It provides a comparable upper
limit to BaBar~\cite{darkPhoton_BABAR1, darkPhoton_BABAR_new} in the
studied $m_{\gamma'}$ mass range. 
Also shown are the exclusion
limits from KLOE~\cite{darkPhoton_KLOE1,darkPhoton_KLOE2,darkPhoton_KLOE3,darkPhoton_KLOE4},
WASA-at-COSY~\cite{darkPhoton_WASA}, HADES~\cite{darkPhoton_HADES},
PHENIX~\cite{darkPhoton_PHENIX}, A1 at
MAMI~\cite{darkPhoton_MAMI2,darkPhoton_MAMI},
NA48/2~\cite{darkPhoton_NA48}, APEX~\cite{darkPhoton_APEX}, and the
beam-dump experiments E774~\cite{darkPhoton_E774}, and
E141~\cite{darkPhoton_E141}.
The $\varepsilon$ values, which would explain the discrepancy between
the measurement and the SM calculation of the anomalous magnetic
moment of the muon~\cite{Pospelov} are displayed in Fig.~\ref{result} as
the bold solid line with a 2$\sigma$ band.\\  

In conclusion, we perform a search for a dark photon in the mass range
between 1.5 and 3.4~GeV/$c^{2}$, where we do not observe a significant
signal. We set upper limits on the mixing parameter $\varepsilon$
between 10$^{-3}$ and 10$^{-4}$ as a function of the dark photon mass
with a confidence level of 90\%. This is a competitive limit in this
dark photon mass range. The BESIII results, which are based on two years
of data taking, are already competitive to the large BaBar data
samples, based on 9 years of running. This is possible due to the use
of untagged ISR events for the dark photon search as well as 
the fact that the center-of-mass energy of the BEPCII collider
is closer to the mass region tested. We also use
a different analysis approach, which has no dependence on the radiator function.\\


The BESIII collaboration thanks the staff of BEPCII and the IHEP computing center for their strong support. This work is supported in part by National Key Basic Research Program of China under Contract No. 2015CB856700; National Natural Science Foundation of China (NSFC) under Contracts Nos. 11235011, 11335008, 11425524, 11625523, 11635010; the Chinese Academy of Sciences (CAS) Large-Scale Scientific Facility Program; the CAS Center for Excellence in Particle Physics (CCEPP); Joint Large-Scale Scientific Facility Funds of the NSFC and CAS under Contracts Nos. U1332201, U1532257, U1532258; CAS under Contracts Nos. KJCX2-YW-N29, KJCX2-YW-N45, QYZDJ-SSW-SLH003; 100 Talents Program of CAS; National 1000 Talents Program of China; INPAC and Shanghai Key Laboratory for Particle Physics and Cosmology; German Research Foundation DFG under Contracts Nos. Collaborative Research Center CRC 1044, FOR 2359; Istituto Nazionale di Fisica Nucleare, Italy; Joint Large-Scale Scientific Facility Funds of the NSFC and CAS; Koninklijke Nederlandse Akademie van Wetenschappen (KNAW) under Contract No. 530-4CDP03; Ministry of Development of Turkey under Contract No. DPT2006K-120470; National Natural Science Foundation of China (NSFC); National Science and Technology fund; The Swedish Resarch Council; U. S. Department of Energy under Contracts Nos. DE-FG02-05ER41374, DE-SC-0010118, DE-SC-0010504, DE-SC-0012069; University of Groningen (RuG) and the Helmholtzzentrum fuer Schwerionenforschung GmbH (GSI), Darmstadt; WCU Program of National Research Foundation of Korea under Contract No. R32-2008-000-10155-0.





\end{multicols}
\end{document}